\documentclass[journal]{IEEEtran}
\ifCLASSINFOpdf
\else
\fi

\usepackage{graphicx}
\usepackage{multirow}
\usepackage{wrapfig}
\usepackage{amsmath}
\usepackage[]{algorithm2e}
\usepackage{algorithmic}
\usepackage{xcolor}
\usepackage{color,soul}
\usepackage{comment}
\usepackage{todonotes}
\usepackage[caption=false]{subfig}
\usepackage{url}

\begin{document}
%
\title{Play with One's Feelings: A Study on Emotion Awareness for Player Experience}
%
%
%

\author{Yoones A. Sekhavat,~\IEEEmembership{Member,~IEEE},
	Samad Roohi,~\IEEEmembership{Member,~IEEE},
	Hesam Sakian Mohammadi,~\IEEEmembership{Member,~IEEE}
	and Georgios N. Yannakakis,~\IEEEmembership{Senior Member,~IEEE}
\thanks{Dr. Yoones A. Sekhavat, Samad Roohi and Hesam Sakian Mohammadi are with the Faculty of Multimedia, Tabriz Islamic Art University, Azadi Blvd, Tabriz, Iran, 5164736931, e-mail: sekhavat@tabriziau.ac.ir}
\thanks{Georgios N. Yannakakis is with the Institute of Digital Games, University of Malta, Msida, MSD 2080, Malta and the School of Electrical and Computer Engineering, Technical University of Crete, Chania, GR73100, Greece.}
}

%
%

\markboth{IEEE Transactions on Games}%
{}
%



\maketitle

\begin{abstract}
Affective interaction between players of video games can elicit rich and varying patterns of emotions. In multiplayer activities that take place in a common space (such as sports and board games), players are generally aware of the emotions of their teammates or opponents as they can directly observe their behavioral patterns, facial expressions, head pose, body stance and so on. Players of online video games, however, are not generally aware of the other players' emotions given the limited channels of direct interaction among them (e.g. via emojis or chat boxes). It also turns out that the impact of real-time emotion-awareness on play is still unexplored in the space of online digital games. Motivated by this lack of empirical knowledge on the role of the affect of others to one's gameplay performance in this paper we investigate the degrees to which the expression of manifested emotions of an opponent can affect the emotions of the player and consequently his gameplay behavior. In this initial study, we test our hypothesis on a two-player adversarial car racing game. We perform a comprehensive user study to evaluate the emotions, behaviors, and attitudes of players in \emph{emotion aware} versus \emph{emotion agnostic} game versions. Our findings suggest that expressing the emotional state of the opponent through an emoji in real-time affects the emotional state and behavior of players that can consequently affect their playing experience.
\end{abstract}


%
\IEEEpeerreviewmaketitle

\section{Introduction}\label{sec:introduction}
	
    Digital games are designed to offer rich affective experiences for players through their interactions with the various forms of game content and types of non-player or player agents. Multi-player digital games extend the affective capacities of a game as they make it possible to play against a number of different opponents (adversarial play) or teammates (collaborative play) including friends, strangers or artificial intelligence (AI)-controlled agents~\cite{yannakakis2018artificial,gaina20172016,schiller2018inside}. Such social interactions between players in multi-player games may result in augmented involvement and immersion during gameplay~\cite{ravaja2006spatial,yannakakis2011experience,sekhavat2017comparison,sekhavat2018sense}. One can only assume that the understanding of such experiences and the study of player emotion is both crucial and beneficial to modern game design~\cite{yannakakis2014guest}. The relationship between emotion awareness, however, and in-game experience and behavior is currently understudied. 
	
    Humans are social beings with motivations for social interaction. Given that social relationships are affective by nature, we expect that playing against opponents that manifest varying emotions can elicit different emotional responses to the player. As emotions in social settings can affect human decision making, we also expect that the awareness of the opponent's emotion may have an influence on the behaviors of a player. Motivated by the lack of a comprehensive empirical study examining the impact of emotion awareness in multi-player settings, this paper investigates the influence of an opponent's emotions on the player's state. As a first step, this initial study examines the degrees to which a player's emotional and behavioral state is affected by manifested emotional expressions of an opponent---via displayed emojis---in a two-player car racing game. Although earlier research has studied the emotional impact of playing against different player types~\cite{cagiltay2015effect,ravaja2006spatial}, to the best of our knowledge, no other study has examined the effect of emotional awareness of opponents to the player's behavior and experience. In particular, we test the following hypotheses within the genre of two-player adversarial car racing games:
	
	\begin{description}
		\item[\textbf{H1}] The emotional state of the opponent (as expressed through an emoji) attracts the attention of the player.
		\item[\textbf{H2}] The behavior of a player is affected by the emotional state of the opponent.
		\item[\textbf{H3}] The emotional state of a player is affected by the emotional state of the opponent.
		\item[\textbf{H4}] The awareness of the opponent's emotional state enhances the perceived competitiveness.
		\item[\textbf{H5}] The awareness of the opponent's emotional state increases the sense of social presence.
		\item[\textbf{H6}] The awareness of the opponent's emotional state increases the player's reported enjoyment.
	\end{description}
	
    To test the aforementioned hypotheses, we developed a special-purpose two-player car racing game equipped with a eye-tracking device which captures gaze data of players, and a facial expression analysis component that detects the emotional state of the player during play. We designed two game modes and conducted a game user study with 20 participants. Players were asked to play the game in the \emph{emotion aware} mode---in which the emotional state of the opponent is presented to the player during play---and in the \emph{emotion agnostic} mode---in which the player is not aware of the opponent's emotions. Experiments in this paper test collectively how the awareness of the emotional state of an opponent affects the emotions and consequently the gameplay behaviors of players. It is important to note that these are facial expressions that are naturally manifested as a result of playing a game and not deliberately self-reported. While the self-report of emotions via emojis is a popular game design practice---as e.g. in \emph{Clash Royale} (Supercell, 2016)---the real-time detection of natural expressions proposed here offers more flexibility and granularity for the design of games and their experience. Self-reported emojis in games could form a valid experimental control against manifested emojis during gameplay; such an experiment, however, is outside the scope of this paper. 
	
    Our key findings suggest that all but one (H5) hypotheses are validated indicating that the expression of the opponent's emotional state through an emoji affects the emotional and behavioral state of players. Finally, it seems that being aware of an opponent's affective state yields higher perceived competition and enjoyment.

\subsection{Why Facial Expression?}

    A number of methods can be used to detect and classify the emotional state of a player from signal streams that are manifested during the game. Among the various modalities of player input that are available, we focus on the analysis of a player's facial expression as it offers a number of key advantages. First, player experience can be inferred automatically based on facial expression detection and affect modeling analysis. Second, various degrees of in-game challenges can also be determined from facial expressions~\cite{blom2014towards}. Finally, emotion detection via facial cues is a non-intrusive task that can be performed using simple off-the-shelf webcameras~\cite{burns2017detecting}. In this paper we use facial expression analysis techniques to detect and analyse the emotional states of racing game players and in particular, we study the \emph{happy}, \emph{sad}, \emph{angry}, or \emph{neutral} player states as manifested through facial expression. 

\subsection{Why Car Racing Games?}
	
    All of our hypotheses regarding the expression of manifested emotions of opponents are tested in a two-player adversarial car racing game. There are three reasons for selecting this game genre. First, driving in common racing games is not conducive to rich social interactions between players because players can only see the cars; normally players are not able to see the drivers and their emotional expressions. This common design feature of the racing game genre suppresses the social presence of another human player behind the steering wheel; hence, racing games can directly benefit from the expression of manifested emotion as the approach followed in this study can enhance the emotional relations between car racing players. Second, the task of playing car racing games well is challenging to study as it requires substantial spatial coordination and kinaesthetic skills even though the possible actions of a player are somehow simple and limited (i.e. accelerating, braking and steering). Third, unlike planning-heavy games such as strategy games that require the player to wait for a long period before seeing the consequences of their actions, fast-paced racing games offer instant feedback of a player's actions that, in turn, affect the player's performance. 

	\section{Background: Emotion Expression in Games}

	The study of emotions in games is an area with an increasingly central role within game research \cite{yannakakis2018artificial}. The understanding of players' emotions and their linking to playing experience is one of the main targets for game design \cite{yannakakis2014emotion}. In addition, affect-based game interaction can result in emotional patterns which may enhance the playing experience. Emotions in games are realized within an affective loop, in which the game elicits, detects and responds to the emotions of players. Various artificial intelligence algorithms and human-computer interaction methods can be used for such purpose~\cite{yannakakis2018artificial,yannakakis2014emotion}. When it comes to racing games and the investigation of player experience in that genre Tognetti et al. \cite{tognetti2010enjoyment} have developed a framework that detects the levels of player enjoyment directly from physiological signals  via the use of preference learning. Georgiou and Demiris. \cite{georgiou2017adaptive} took a further step forward for the detection of a car racing player state by integrating additional modalities of user input to the player model.
	
	Earlier studies have shown that the core characteristics of an opponent may have an impact on the affective or cognitive state of a player of a competitive computer game~\cite{weibel2008playing,mandryk2006using}. For example, different physiological responses can be observed in players when playing against a computer or a friend. In particular, playing against another person elicits greater arousal compared to playing against a computer opponent~\cite{weibel2008playing}. It has also been shown that playing with friends can result in higher degrees of spatial presence, engagement, and physiological arousal than playing with strangers~\cite{cagiltay2015effect}. Mandryk et al. has also reported different physiological responses of players when playing against a computer or a friend~\cite{mandryk2006using}. As discussed in~\cite{jervcic2018effect}, human collaborators are perceived as more credible and socially present than non-human collaborators in the context of serious games. On the other hand, rich social interactions can be established within temporary teams in games, as players are interested in collaborating with strangers~\cite{kou2014playing}. 
	
    Another important element that can affect the behavior of players in the social context of games is competition~\cite{dondlinger2007educational}. In competition-based (adversarial) games, one player advances toward achieving a goal, while the other moves further away from it~\cite{hong2009playfulness}.  Prior research has shown the potential of competition to draw the attention and excitement of players in games~\cite{cheng2009equal} that can, in turn, motivate players to put more effort into the in-game tasks. Playing a competitive video game activates the cognition associated with motivational pathways~\cite{katsyri2013just}. Research has shown that competitive video gaming can lead to enjoyment and positive affect~\cite{vorderer2003explaining}. Competition is also an important element of a computer game that can affect the motivation of players~\cite{dondlinger2007educational,melhart2019your}. Research has shown that although competition is not an essential game design element, it is nevertheless a critical motivator for gameplay experience and immersion~\cite{shaffer2006computer}. In this study, we argue that the awareness of the emojis of game opponents can affect the competition levels among players that can consequently lead to more enjoyment and positive affect in the game. In particular, as examined through hypothesis H4, we argue that the awareness of the opponent's emotional state enhances the perceived competitiveness.
	
	Research has shown that emotional expressions can convey a person's cooperative tendencies that can, in turn, affect the decision-making of others~\cite{hoegen2017incorporating}. For example, smiles can indicate cooperative tendencies~\cite{stratou2015emotional} and temporal dynamics of smiles can shape trust among players of games~\cite{krumhuber2007facial}. Inexpressive opponents, instead, are generally viewed as untrustworthy~\cite{schug2010emotional}. Following this line of research Hoegen et al.~\cite{hoegen2017incorporating} have built a model that predict a player's behavioral patterns based on the opponent's emotional expressions and actions; such a model can be used for planning social interactions within the game. Decoding emotional signals of opponents has also been used in negotiation. According to~\cite{de2014reading}, people can make inferences about others' mental states from emotional expressions for decision making during a negotiation phase; it appears that facial expression contains critical cues for making such inferences.  Further research in behavioral science has studied the impact of emotions on one's own behavior and on another's behavior in negotiations~\cite{van2010interpersonal}. For example, it is shown that manifesting anger during a negotiation can yield more concessive behaviors on one's opponent~\cite{pietroni2008emotions}, whereas expressing happiness may lead to fewer concessions. Emotions can further be used for the purpose of resolving conflicts in agent-agent or human-agent interactions~\cite{yannakakis2010siren}. Van Kleef et al. contend that ``each discrete emotion has its own antecedents, appraisal components, relational themes, and action tendencies''~\cite{van2010interpersonal}. They argue that observing a particular emotion in a person provides ``relatively differentiated information about how that person regards the situation''~\cite{van2010interpersonal}. Considering video gaming as a social interaction, our findings in this paper highlight that expressing emotion plays a core role in controlling the behaviors of opponents in this game-based form of social interaction.
	
	\section{Experiment Design}
	
	To test our core hypothesis that opponent emotion awareness has an impact on the player experience we conducted a user study in a controlled environment (computer lab) and invited 20 participants to play variants of a two-player car racing game. The details of the experimental protocol we used and the game user study are detailed in the remainder of this section.
	
	\subsection{Car Racing Game}
	
		\begin{figure}[!tb]
		\centering
		\includegraphics[width=3.5in]{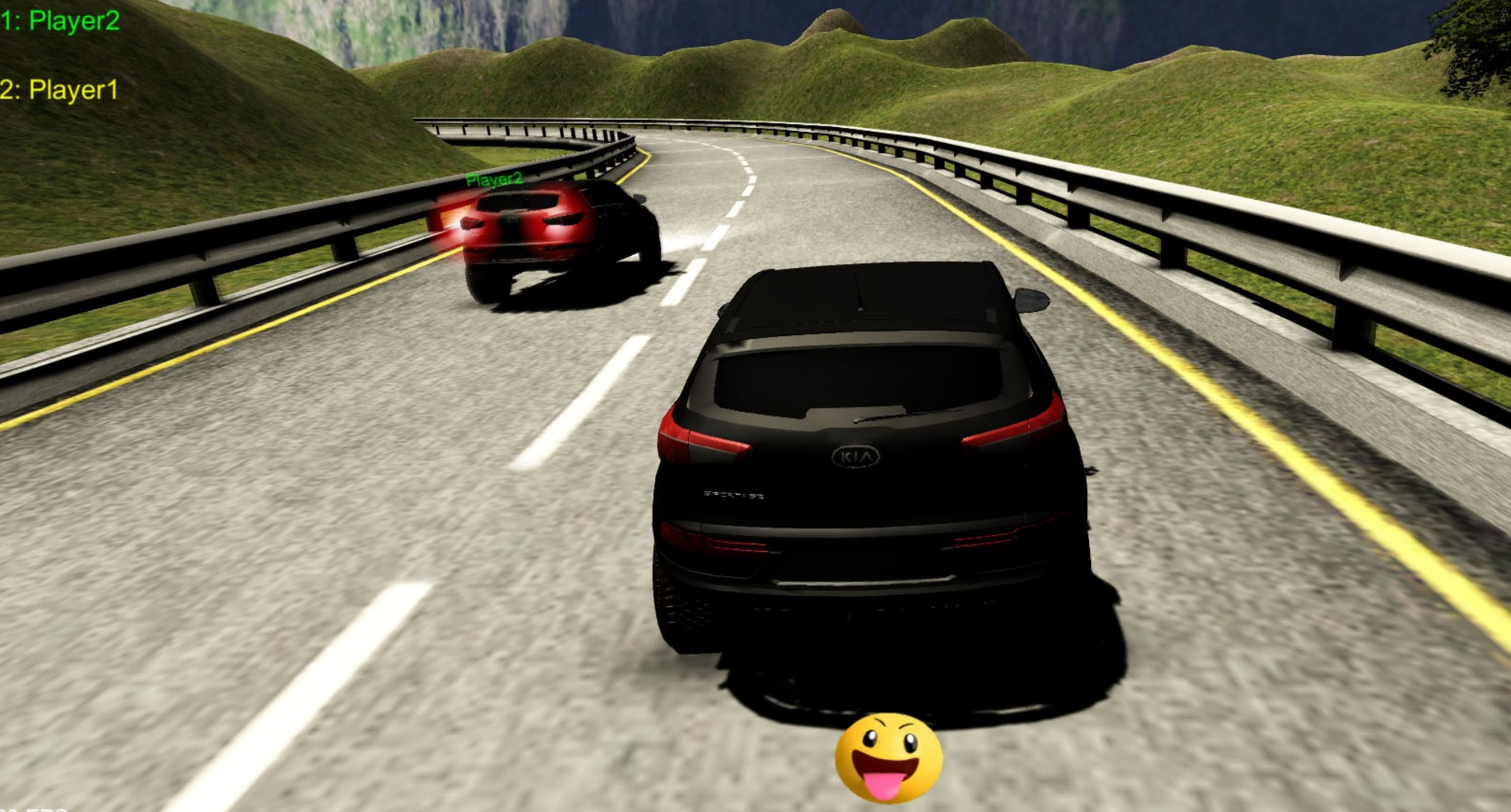}
		\caption{A screenshot of car racing game developed with the presentation of an emoji at the bottom-center of the player's car.}
		\label{Fig-Game}
	\end{figure}

	A prototype of a car racing game was developed and used in all experiments reported in this paper; see Fig. \ref{Fig-Game}. In this game, a player can control the car by accelerating, braking, and steering to right and left. This game provides adversarial gameplay through which two or more players can compete against each other to gain the first place in a racing route. After the initialization phase, the main loop of the game is started, where user inputs are gathered from the input devices (gaze and facial expression tracker) that are described below. 
	
	\subsection{Emotional Tracking via Facial Expression}
	
	This paper studies the relationship between a player's and an opponent's manifested emotions as displayed during the game. The particular set of emotional states chosen for this study is based on Ekman's taxonomy of basic emotions as expressed through faces~\cite{ekman1999basic}. In this study, we focus on \emph{happiness}, \emph{sadness}, and \emph{anger} as \emph{disgust} and \emph{fear} are not relevant emotions for this game genre and \emph{surprise} (even though highly relevant for the racing genre) was not detected as frequently by our facial expression detection software and thus it was not included in our analysis.
		
	We captured a player's emotions through a facial expression recognition system consisting of a face detection module, a preprocessing module, a deep neural network module, and a mapping module. In particular, we use a supervised transform network~\cite{chen2016supervised} as a face detector. The captured face image is preprocessed to meet the constraints of a fine-tuned version of VGG-16 \cite{simonyan2014very} which is used, in turn, to compute a player's emotion with 13 convolutional layers, and 3 fully connected layers. To retrain the network for the purpose of facial expression recognition, we used the Extended Cohn-Kande dataset (CK+)~\cite{lucey2010extended}, which provides a probability distribution over seven distinct emotional states, including neutral, anger, happiness, fear, sadness, disgust and surprise. The results of our training experiments yield a test accuracy of $94.98\%$ on the CK+ dataset~\cite{yang2017facial}\footnote{The source code of our face recognition system is freely available at \url{http://www.carlab.ir/}.}
	
	\subsection{Participants}
	
	For this study we recruited 20 participants (16 males and 4 females) from undergraduate and graduate students of the faculty of Multimedia at Tabriz Art University. Participants' ages range between 21 and 27 years old, with a mean of $24.5$ years. Upon recruitment, participants completed a brief questionnaire regarding their age, gender, video game skill level and time spent regularly playing. Those who reported no experience with video games were excluded. According to the background information captured through the pre-study questionnaire, all of the participants were frequent gamers. Based on the classification provided in~\cite{green2003action}, frequent gamers are those who play video games at least 5 hours a week for a period of 6 months or more continuously. Subjects participated in the experiment in groups of two persons of the same sex.
	
	\subsection{Experimental Protocol}
	
	To better distinguish between the settings with and without emotional awareness in a car racing game, we designed a within-subjects user study, where each participant played the game in both settings. In the emotion aware setting (EAW), the facial expression of each player is identified based on the image captured from the webcam installed above the monitor. Then, the emotional state based on this facial expression is presented to the opponent as expressed through the facial expression of an emoji. In our attempt to find the best possible position to place the emojis on the screen, we decided to display them just below the player's car---which is always at the bottom-center of the screen---for two reasons. First, as the player always looks at her car while playing, the emojis are always visible in the area close to the player's gaze. Second, there is a good contrast between the emoji and the solid color of the background (road) that makes it easy to recognize the emoji without additional cognitive burden (see Fig. \ref{Fig-Game}). In the emotion agnostic (EAG) setting while the facial expression of opponents is extracted and logged, this information is not displayed to the players. 

\begin{figure}[!tb]
		\centering
		\includegraphics[width=2.0in]{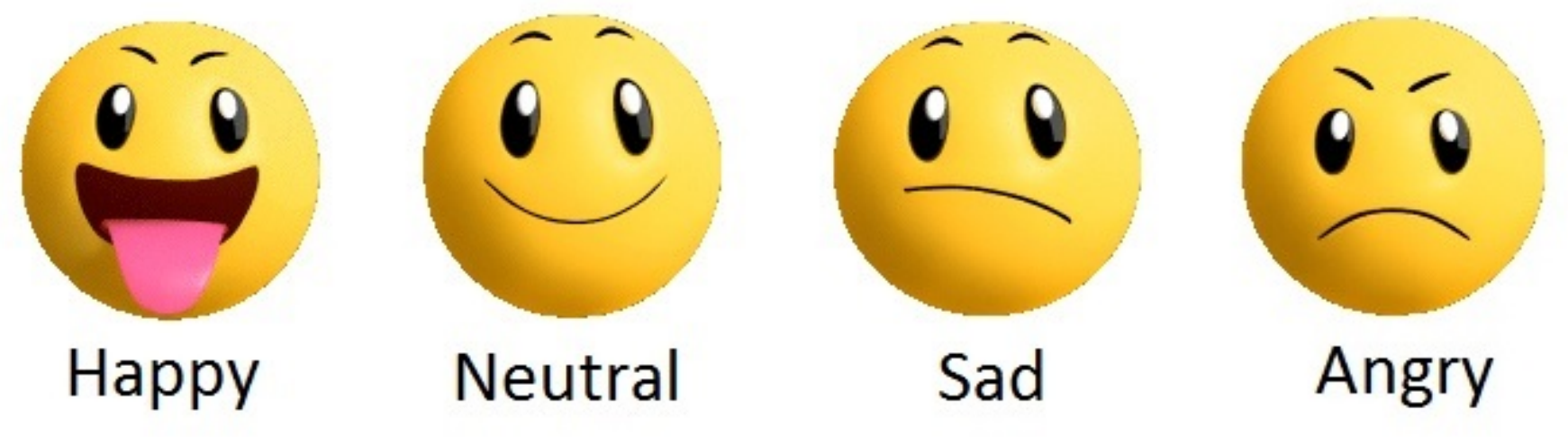}
		\caption{The emojis used to express different emotions of the opponent in the game.}
		\label{Fig-Emotions}
	\end{figure}
	
     The emojis used in the prototypes were downloaded from vexels\footnote{Available at \url{https://www.vexels.com/graphics/emoji}}. Based on these images, we created 2-second animations of facial expressions representing the transitions from one emotional state to another. The final frames of different emotional states are shown in Fig.~\ref{Fig-Emotions}. As illustrated in that figure we represent the happiness state by using a grinning face with smiling eyes; we used a slightly smiling face with neutral eyes to represent the neutral state.
	
	Each emotional state was accompanied by a corresponding short sound effect. By using auditory effects in conjunction to the emojis we wish to ensure that the player is aware of the current emotional states of her opponent. The ambient background sound of the car racing game is a typical sound of a car engine. A change in the emotional state of the opponent was a trigger for playing the corresponding sound as well as for playing the transition animation (a change from one emoji to another). We selected short sounds (around 2 seconds) for each emotion among the tagged sounds available in zapsplat\footnote{\url{https://www.zapsplat.com}}. To avoid frequent and abrupt changes in the presentation of emotional states, we skipped the changes that were happening during the transition from one state to another.
	
	In order to address potential order and learning effects, the order of exposure to the different settings was varied, resulting in the assignment of participants to two groups: the first group started the experiment with the EAW setting whereas the second group started playing on the EAG setting. This way, the order in which the participants play in different settings was eliminated as an independent variable.
	
	Players answered a set of questions before and after the experiment. Before playing the game, each player was informed about the controls and the user interface of the game including the facial expression of the opponent (in the EAW mode). As discussed in~\cite{backlund2006computer}, players who are not assigned any specific task during a driving game enjoy the game more than the players who were given a task. Consequently, we let players to freely drive their car, and play and enjoy the game in their own way. In groups of two players, each player is asked to drive a customized track for three laps. The first lap is a training lap, in which we did not measure the performance data and players had the opportunity to become familiar with the controls of the game. The driving path was specifically designed for this study and players did not have an experience of driving on this route before the study. After playing the game, participants were asked to answer some informal questions regarding the game. 
	
	\begin{figure}[!tb]
		\centering
		\includegraphics[width=3.5in]{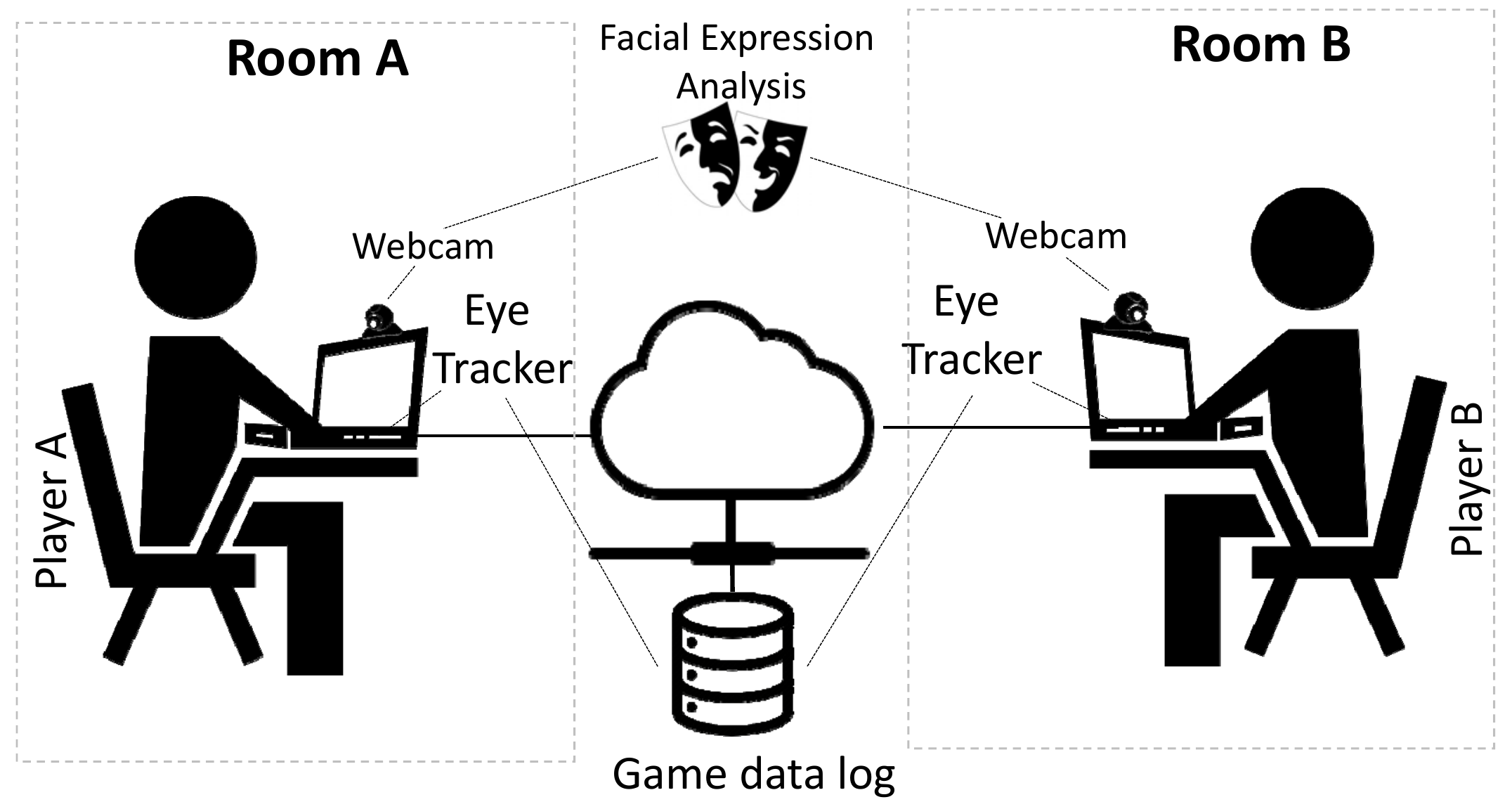}
		\caption{The overall setting of the user study. The facial expression module as well as the eye-tracking device is active in EAW and EAG modes. The emotional state of the opponent is shown to the player only in the EAW mode.}
		\label{Fig-TestEnvironment}
	\end{figure}
	
	\begin{figure}[!tb]
		\centering
		\includegraphics[width=3.5in]{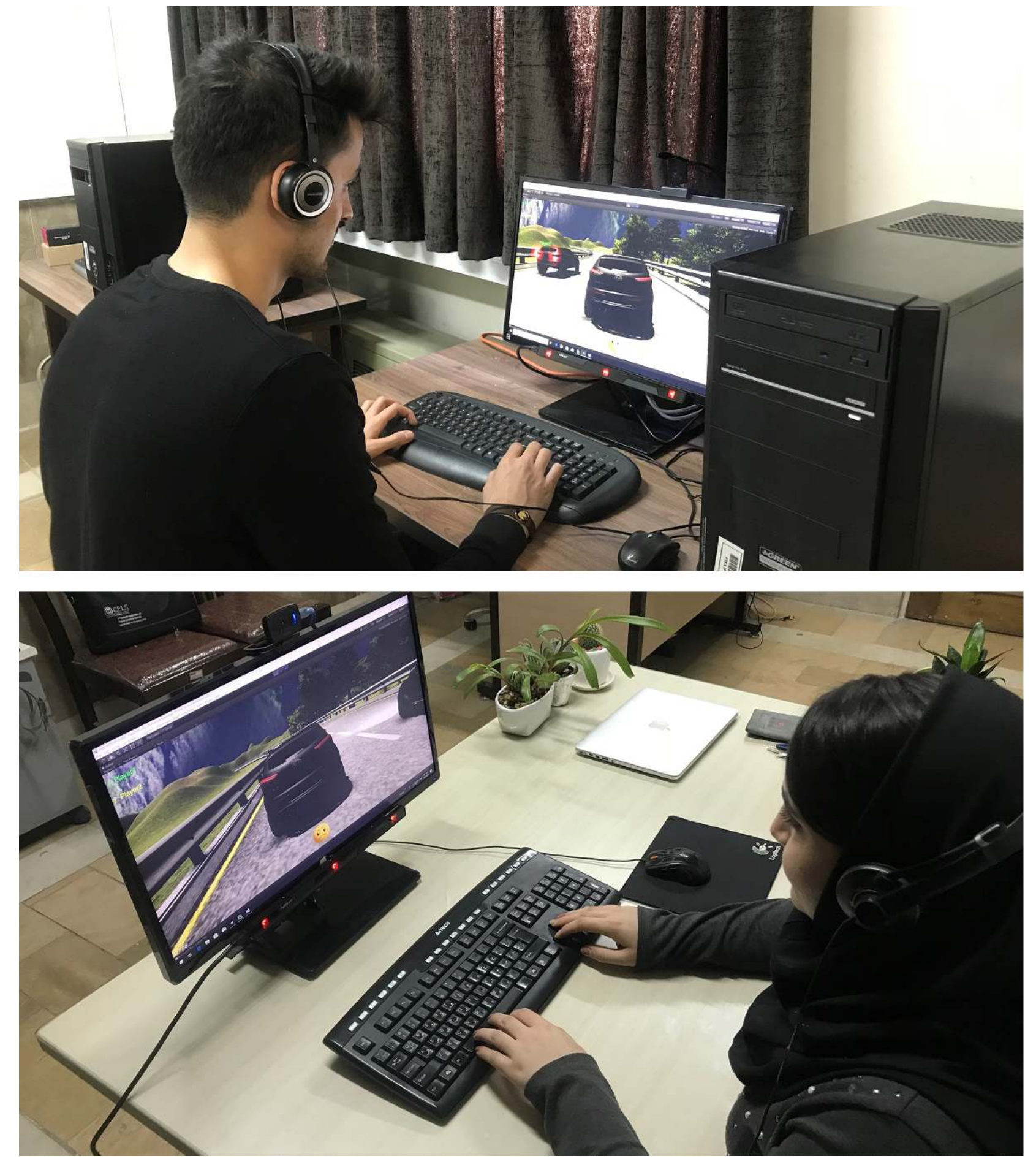}
		\caption{Two players in separate rooms who are competing against each other in the car racing game. Webcams and eye trackers are placed, respectively, above and below the monitor.}
		\label{Fig-Playing}
	\end{figure}
	
	As shown in Fig.~\ref{Fig-TestEnvironment} and Fig.~\ref{Fig-Playing}, participants were asked to sit in front of a 24-inch monitor and play the car racing game with a standard keyboard. As the attachment of sensors to body parts is an intrusive process that may affect the experience of players \cite{yannakakis2018artificial,yannakakis2014emotion} we used an unobtrusive eye tracker to capture gaze data. In particular, we installed a Tobii EyeX eye tracker to capture the eye movements of players and their focus on the screen during driving. Data from the game including speed, position, number of collisions, players' actions (braking, speeding or steering), facial expression as well as eye gaze data was all collected during the game and time-stamped. Real-time performance metrics including speed and the position of the car in the racing track, as well as the facial expression of the opponent, were displayed on the monitor.
	
	Before starting the experiments---when participants had an opportunity to freely play the game and become familiar with its controls---we explicitly informed participants that the emojis on the screen represent the real-time facial expression of their opponent while playing. This way, we made sure that players attribute the emoji to the opponent's emotion and not to their own.

	\subsection{Measures}
	
	To test each one of our five hypotheses we relied on a number of measurable parameters that are detailed in this section.
	
	\subsubsection{Attention to Emotional Expression}
	
	In addition to using gaze as a sole modality to control the game~\cite{munoz2011towards}, eye-tracking technology can offer information on where and what players are looking at while playing. Since in the EAW mode of the game the emotional state of the opponent is shown as a face icon on the screen, one would expect that players would pay attention to the icon during the game if it is relevant for their gameplay experience. As mentioned earlier we used two eye-tracker devices (one for each player) to obtain gaze data on the screen. The analysis of this data could reveal if the emotional states expressed via emojis are under the player's attention.
	
	\subsubsection{Player Behavioral and Performance Metrics}
	
	Since the aim of this paper is to study the degree to which the emotional state of one player may affect the behavior and the emotion of the other player, we capture a number of metrics regarding the behavior of players in the game as well as measures that indicate in-game performance. In particular, we log the position of a player, the speed of the car, the ranking of a player during the race, the state of focusing on the emoji (a binary value indicating if a user looks on the emoji or not in ea), the number of collisions with other cars on the road or some objects on the roadside (as an indicator of driving mistakes) and the emotional state of the player.  
	
	\subsubsection{Self-Reports}
	
	We measured two subjective factors using self-report questionnaires. First, participants were asked to answer some questions regarding \emph{Perceived Competitiveness}. This parameter was measured by six statements adapted from~\cite{song2013effects} as follows: 1) I felt that this game was competitive; 2) I think other people would feel that this game is competitive; 3) I paid attention to my position in the route in relation to the position of my opponent; 4) I tried to be ranked first; 5) Ranking on the screen motivated me to drive better; 6) Other people's performances motivated me to drive better and try harder. Participants were asked to represent their levels of agreement on a 5-point Likert-type scale. Participants also reported their social presence in the game. To this end, we measured the social presence using 17 questions~\cite{ijsselsteijn2013game} that assess the psychological and behavioral involvement of the player with other social entities. Finally, participants were asked to answer whether presenting the emotions of the opponent through the emojis had a positive effect on their enjoyment while playing.
	
	A summary of the variables measured to support or reject each hypothesis and the corresponding method used to measure the variables are shown in Table~\ref{tbl-Variables}.
	
	\begin{table*}
		\caption{The variables used to test each of the six hypotheses in this paper and the corresponding method used to measure each variable}
		\label{tbl-Variables}
		\begin{center}
			\begin{tabular}{ | p{5cm} || p{5.7cm} | p{5.7cm} |}
				\hline\hline
				Hypothesis & Variables & Method\\ \hline\hline
				H1: Attracting the attention of players. & Eye gaze data, fixation count, dwell time. & Data collected by Tobii EyeX eye-trackers.\\ \hline
				H2: Effect on the behavior of players. & The number of braking, steering and throttling, the number of collisions, gameplay duration. & Logged keyboard inputs. Captured in-game events and properties such as collisions with other cars and objects on the road, and racing times.\\ \hline
				H3: Effect on the emotional states of players. & The number of happy frames and angry frames, the number of emotionally active states, 
				 the rate of changes in the emotional states of players. & Using a facial expression detection algorithm to detect emotions, finding relations between the emotional states of players and opponents, and comparing emotionally active states across different modes. \\ \hline
				H4: Enhancing the perceived competitiveness. & Perceived competitiveness. & Six
				questions adapted from the perceived competitiveness questionnaire~\cite{song2013effects}. \\ \hline
				H5: Increasing the sense of social presence. & Social presence. & 17 questions based on social presence questionnaire ~\cite{ijsselsteijn2013game}. \\ \hline
				H6: Increasing reported enjoyment. & Reported enjoyment. & A single question about the effect of presenting the emotions of the opponent on game experience.\\ \hline\hline
			\end{tabular}
		\end{center}
	\end{table*}
	
	\section{Results}
	
	The collected quantitative and qualitative data was analyzed and the results of this analysis are presented in the following sections; each section is dedicated to each of the five hypotheses of this paper. 
	
	\subsection{Attention to Emotional Expression: H1}
	
	Before studying the effect of opponent emotional awareness on the behavioral and emotional states of players, we test the degree to which players pay attention to the emotional state of opponents displayed for them as an emoji at the bottom-center of their game screen. To this end, we used the data captured from the two Tobii EyeX eye-trackers. 
	
	A point captured by an eye-tracker is a single gaze that has a specific position. Gaze points close together can be grouped to form a fixation. Research has shown that fixations (i.e., moments when the eyes are relatively stationary), can reveal which parts of the displayed information are most salient, which is linked to attention~\cite{eckstein2017beyond}. Although a fixation does not necessarily mean that the participant perceived an element, generally speaking, elements that draw visual attention have a higher chance of being perceived~\cite{poole2006eye}. In our experiments, we computed fixations as areas on the screen where participants look for at least 200 ms following the guidelines proposed in~\cite{salvucci2000identifying}. Based on~\cite{hornof2002cleaning} we set the maximum distance that a point may vary from the average fixation point and still be considered part of the fixation at 50 pixels. 
	
\begin{figure}[!tb]
		\centering
		\includegraphics[width=0.49\linewidth]{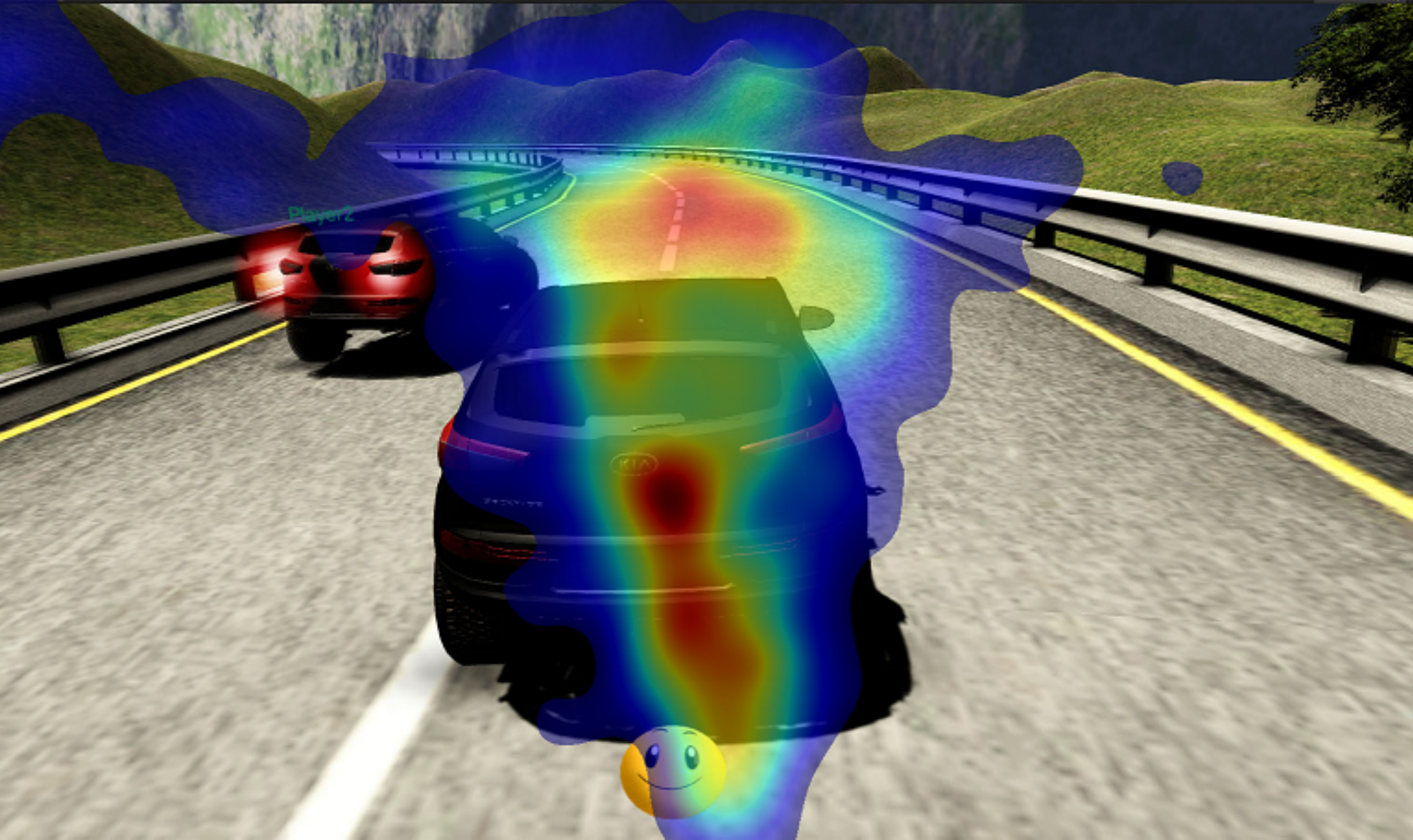}\label{Fig-HeatMap-EA}
		\includegraphics[width=0.49\linewidth]{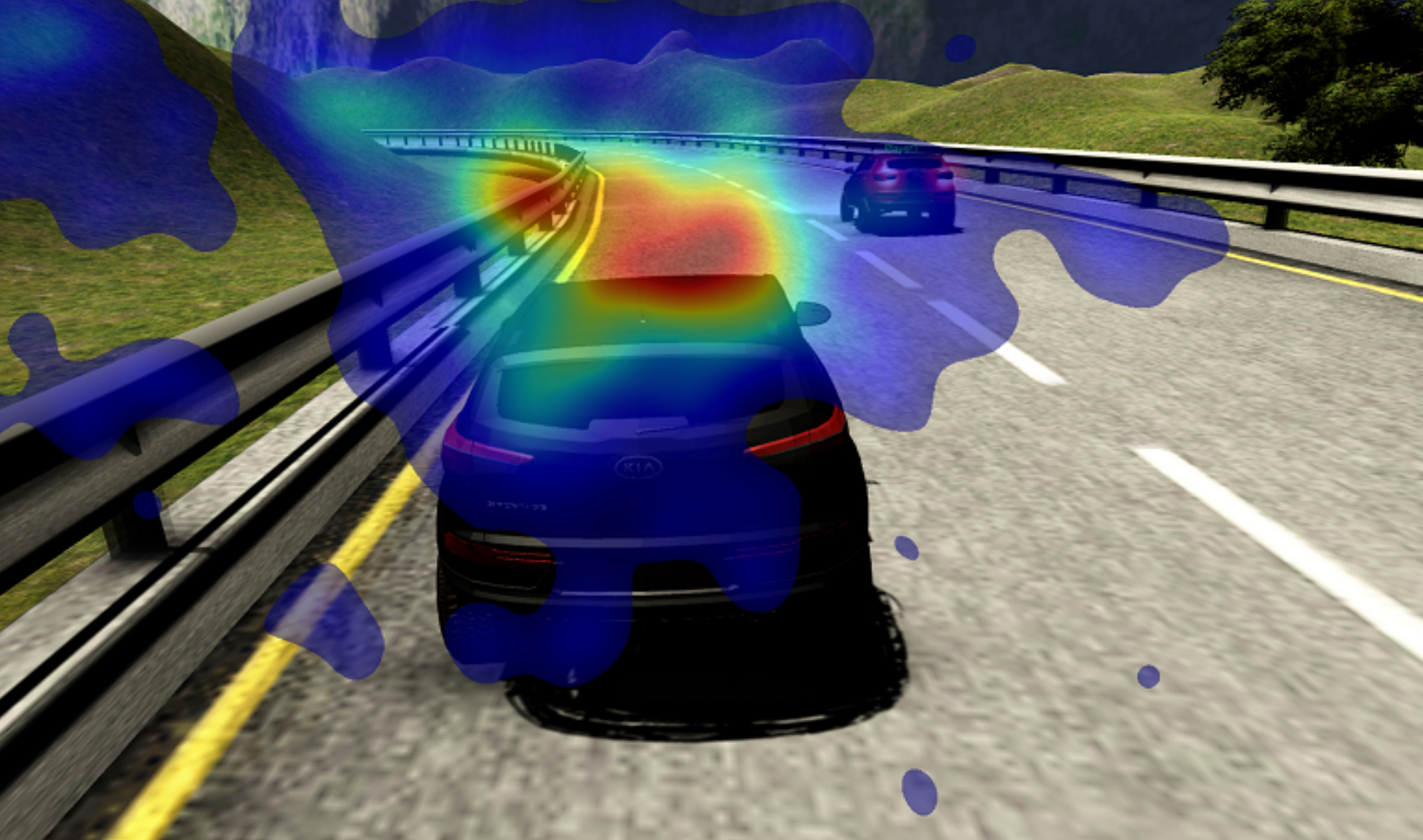}\label{Fig-HeatMap-NEA}
		
		\caption{Heatmaps of eye fixations aggregated across all participants in the EAW mode (left) versus the EAG mode (right). The hotter (red) the area of the map the more gaze fixations are available in that area.} 
		\label{fig:heatmaps}
	\end{figure}

\begin{figure}[!tb]
		\centering
		\includegraphics[width=3.5in]{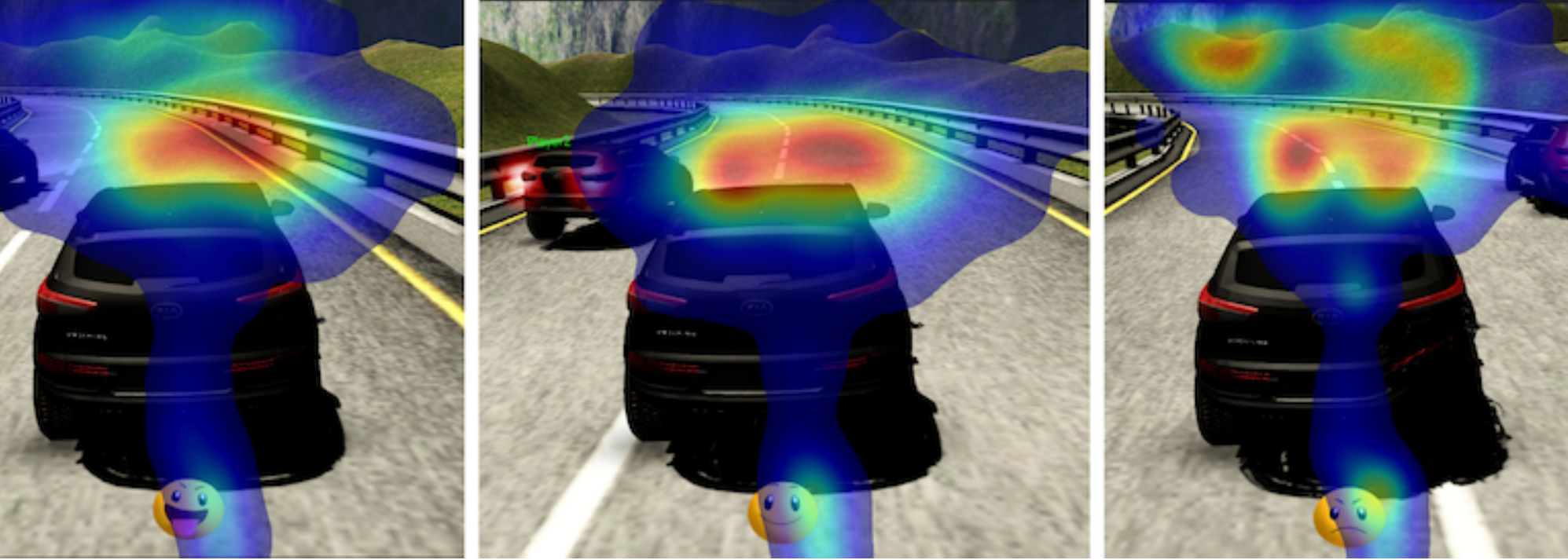}
		\caption{Heatmap of accumulative gaze data across the three different emotional states examined in this work: happy (left), neutral (center), and angry (right).}
		\label{Fig-HeatMap-HNA}
	\end{figure}

	Heatmaps are time-aggregated density-based representations that are often used to visualize gameplay data \cite{yannakakis2018artificial}. When applied to gaze data of players, heatmaps can reveal the areas where the players look at more frequently~\cite{raschke2014visual}. In our case study a heatmap of player fixations may offer the first insights on whether the player pays attention to the emotional awareness of his opponent while playing the game or not. For that purpose the heatmaps or eye fixation data for all players in both EAW and EAG states are visualized and are shown in Fig.~\ref{Fig-HeatMap-EA} and Fig.~\ref{Fig-HeatMap-NEA}, respectively. As it can be observed from these figures the emotional states of the opponent attract the attention of the player only in the EAW mode. Further, the fine-grained analysis of gaze data based on the three dissimilar emotional states of the opponent (see Fig.~\ref{Fig-HeatMap-HNA}) shows that the emoji is always under attention by players regardless of the emotional state of the opponent. The results validate hypothesis H1 that the emotional state of the opponent as expressed by an emoji attracts the attention of the player in two-player car racing games. 
	
	\subsection{Behavior Analysis: H2} 
	
	\begin{figure}[!tb]
		\centering
		\includegraphics[width=3.5in]{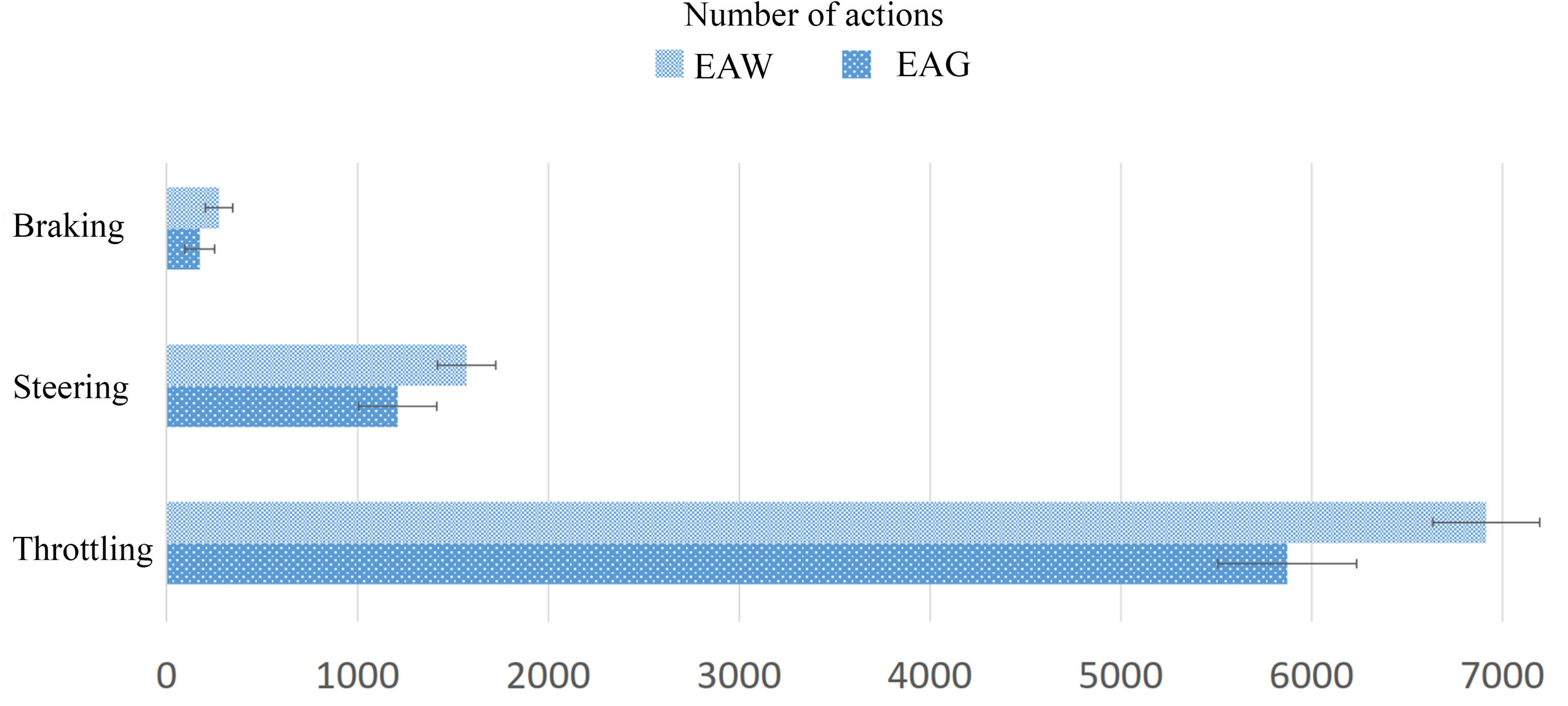} 
		\caption{The average number of braking, steering and throttling across all players in the EAW and EAG game modes. Error bars display the 95\% confidence interval of the average shown.
		}
		\label{Fig-BehavDetails}
	\end{figure}
	
	Through H2 we hypothesize that the emotional state of the opponent---as expressed via emojis---can affect the behavior of the player. Thus, we expect to observe differences in the way the player controls the car in the game in the EAW mode. In the examined game the possible actions a player can take during play include steering (right and left buttons), throttling (up button to speed up) and braking (down button to reduce the speed). Based on these actions we extract a number of core features that characterize the gameplay behavior in a car racing game and examine the degree to which they are affected by the presence of the opponent's emotional manifestations. These include the total amount of steering ($s$), speeding up ($t$) and reducing speed ($b$) but also the total number of collisions with other cars on the road or some objects on the roadside ($n_k$) and the gameplay duration ($t_g$).
	
	As shown in Fig.~\ref{Fig-BehavDetails}, participants were on average more active in terms of steering, braking and throttling when they played the EAW version of the game. To test H2 regarding the effect of one independent variable (EAW vs. EAG) on three dependent variables ($s$, $t$, $b$), we used the repeated measures MANOVA test to cater to and assess multiple response variables simultaneously. The results showed that there was a statistically significant difference in players' behavior based on emotional awareness, $F (3, 23) = 5.1654$, $p < .05$; Wilk's lambda $(\Lambda)$ is $0.622$ ($power=0.828$ according to the post hoc power analysis). Such a finding validates H2 and suggests that the behavior of a player appears to be affected by the emotional state of the opponent in two-player car racing games. 
	
	In addition to car control extracted features, we also compared the behavior of players in terms of the number of collisions with other cars on the road or objects on the roadside, which is an indicator of driving mistakes in the car racing game. We argue that the stress of playing against an emotionally active opponent as well as the distractions coming from the changes in the emoji expressions on the screen can increase the number of these errors. A paired samples t-test was employed to test for any significant difference between the number of collisions in the EAW versus the EAG mode. The result reveals a significant difference between the EAW ($M=11.19, SD=5.75$) and the EAG ($M=7.06, SD=7.49$) modes ($t(19)=2.7139, p=0.0177$, $power = 0.862$) which further indicates that the performance of a player in terms of the number of collisions in the car racing game can be affected by the emotional state of the opponent.
	
	Finally, we compared the EAW and EAG modes in terms of gameplay duration. The gameplay duration is calculated as the total time needed to finish three laps of the path. The results show that participants spend more time in the EAW mode than in the EAG mode. According to the paired samples t-test conducted on the samples, there was a significant difference between EAW ($M=163.34, SD=15.68$) and EAG ($M=134.49, SD=13.48$) modes ($t(19)=2.7436, p=0.01781, power=0.855$) indicating that the emotional awareness of the opponent seems to increase the level of challenge in the game, that can consequently affect the behavior of players. 

	\subsection{Emotion Analysis: H3} 
	
	According to H3, the emotional state of a player is affected by the emotional state of the opponent in two-player car racing games. As mentioned earlier, the facial expression detection algorithm we employ uses the supervised transform network~\cite{chen2016supervised} as a face detector and a VGG-16 network which is further trained on the extended Cohn-Kande dataset (CK+)~\cite{lucey2010extended} to provide a probability distribution over seven distinct emotional states: neutral, anger, happiness, fear, sadness, disgust and surprise. Given the game genre context of car racing, however, only the happy, neutral, sad and angry states appear to be relevant as they cover collectively more than $97\%$ of the predicted emotional states in our dataset; thus, the remaining emotional states were excluded from the analysis of this paper. In addition, as the neutral state is not considered an active emotional state in this study we focus only on the happy, sad and angry states. 
	
	Consequently, we tested the hypotheses regarding the effect of one independent variable (EAW vs. EAG) on two dependent variables ($happy$, $angry$). The counts of frames (60 frames per second) detected as happy and angry states represent the values of the $happy$ and $angry$ variables, respectively. The results of a repeated measures MANOVA on our data suggest that there is a statistically significant difference in players' emotional state based on emotional awareness, $F (3, 24) = 4.580$, $p < .05$; $\Lambda = 0.679$ ($power=0.832$). The results support the hypothesis that the emotional state of the opponent can affect the emotional state of the player.
	
	We also examine in which ways the different emotions of an opponent can affect the emotional state of the player. In particular, how can a happy or an angry opponent affect the emotional state of the player? Although we measure the changes in the facial expression of the player after a change in the facial expression of the opponent, we need to make sure that the facial expression detected on a player derives from looking at the opponent's emoji, and not from other reasons such as encountering different driving situations and losing, or winning the race. To this end, we consider only the facial expression changes that occur after a fixation towards the opponent emoji.
	
	Considering a change in the emoji (representing the emotional state of an opponent) as a stimulus, we expect a change in the emotional state of a player after a reaction time window. Reaction time is the amount of time it takes to respond to a stimulus and varies across different people depending on a number of factors including age, gender, physical fitness, fatigue, distraction, alcohol consumption, personality type, and whether the stimulus is auditory or visual~\cite{jain2015comparative}. According to~\cite{becker2007confounded}, the mean reaction time reported to detect happy and angry faces is around 600 ms. We thus consider 600 ms after a change in the emotional state of the opponent (emoji) as the point in time after which we examine the emotional state of the player. We monitor a time window of 2000 ms after this point within which we detect the emotions of the player. As the emotions can vary during this time window, we consider the dominant emotional state of the player in this period as the emotional state that characterizes it. Given \{happy, sad, angry\} as the set of active emotional states, and \{neutral, noChange\} as the set of inactive states, the changes in the opponent emotional state has resulted in $31\%$ and $19\%$ active states in the EAW and the EAG mode, respectively. We conducted a paired samples t-test to compare the number of active emotional states in the EAW and EAG modes as a change in the expression of the emoji. The result of this test is that there is a significant difference between the active states of the EAW ($M=5.58, SD=1.64$) and the EAG ($M=3.33, SD=1.42$) modes $(t(19)=2.5886, p=0.0224)$, which suggests that players were more emotionally active in the EAW mode than in the EAG mode ($power=0.882$).

\begin{figure}
		\centering
		\includegraphics[width=3.5in]{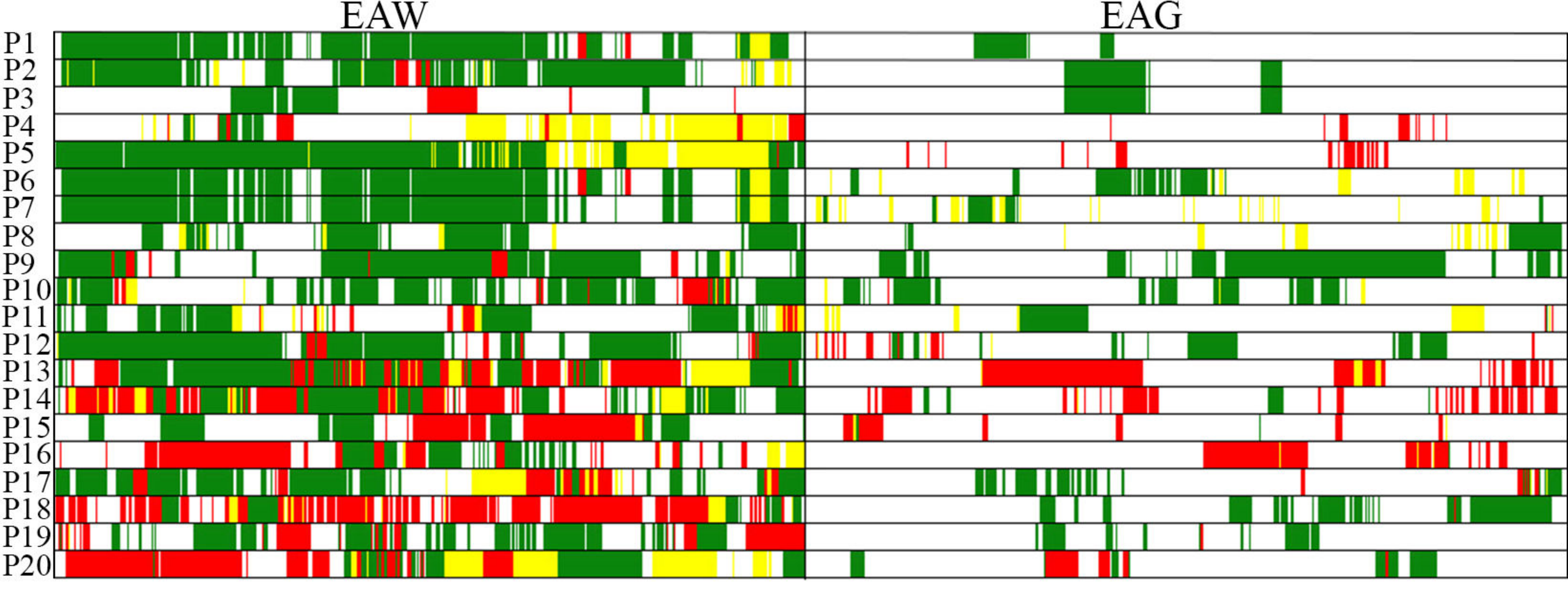}
		\caption{Time sequence of the face-based detected emotions across all players while playing in the EAW and EAG modes. White, green, red and yellow colors illustrate respectively neutral, happy, angry and sad states.}
		\label{Fig-EmotionChanges}
	\end{figure}

	To better visualize the ways players' emotions change over time in both EAW and EAG modes, Fig.~\ref{Fig-EmotionChanges} illustrates the time sequence of the facial expressions of players while playing in both EAW and EAG modes. As shown in this figure, the rate of changes in the emotional states of players in the EAW mode is more salient than in the EAG mode. 
	
	\subsection{Self-Report Measures: H4, H5, and H6}
	
	After playing in both the EAW and the EAG states, participants were asked to complete questionnaires regarding perceived competitiveness and social presence. Because of the relatedness between samples that come from the same population, we used the Wilcoxon Signed Ranks test to highlight statistically significant differences between EAW and EAG conditions in terms of perceived competitiveness and social presence. 
	
	Results from the perceived competitiveness questionnaire showed statistically significant differences between the two tested approaches, suggesting that participants experienced more competition in the EAW than the EAG mode ($Z = -3.441, p = 0.001, power=0.872$). The results validate hypothesis H4 in that the emotional awareness of the opponent can enhance the perceived competitiveness in two-player car racing games. On the other hand, A Wilcoxon signed-rank test on data did not elicit a statistically significant change in the social presence of players ($Z = -1.807, p = 0.071$). Consequently, the hypothesis that the emotional awareness of the opponent increases the sense of social presence in two-player car racing games (H5) is not supported. We argue that this finding can be attributed to the two-player nature of our car racing game, which is inherently social regardless of the degree a player is aware of the opponent's emotions.
	
	Finally, participants were asked to answer if presenting the opponent's emotions through emojis had a positive effect on their enjoyment from the game. As of the within-subjects design of the study, all participants had an opportunity to play the game in both EAW and EAG modes. Consequently, they were in a position to judge if presenting the opponent's emotions through emojis had a positive effect on their gaming experience. Among 20 participants, 16 participants reported positive statements about the presentations of emotions through emojis. In particular, the answers were: ``it was fun to see the emotions of opponents while playing'' (4 persons), ``I liked the emojis, they were motivating'' (2 persons), ``The emojis were cool '' (3 persons), ``I liked it because I could communicate with my opponent through the emojis'' (2 persons), ``It was funny, I think I could affect the emotions of my opponent '' (3 persons) and ``I liked it, even though sometimes during the game I wish I could shut them up'' (2 persons). On the other hand, two participants were not satisfied with the presentation of emotions through emojis. They reported that the presentation of the emojis was annoying and distracting. Two remaining participants reported that their main focus was on the racing and they did not care about the presentation of emotions. This descriptive data validates hypothesis H6 that the awareness of the opponent's emotional state appears to increase the reported enjoyment of players.  
	
	\section{Discussion} 
	
	A core finding of this paper is that the emotional state represented by an emoji seems to attract the attention of players while playing. While arguably gaze fixations do not necessarily correspond to participants perceiving an element on the screen, elements that draw visual attention have a higher chance of being perceived. We also showed that not only the emotional state of an opponent can affect the emotions of the player, but also different emotions may have different effects on the player. Further, our findings showcase that the most frequent emotional state of the player in the EAW mode is the happy state (which is an active state) whereas in the EAG mode, the most frequent state is the neutral state. Consequently, the EAW mode yields scenarios that are more emotionally active than EAG, which can in turn, be used to actively involve the emotions of players in a game. The obtained results also showed that participants were on average more active in terms of steering, braking and throttling in the EAW mode compared to the EAG mode. 
	
	As shown in the results section, the performance of a player in terms of the number of collisions with other cars on the road or objects on the roadside (i.e. an indicator of driving mistakes) as well as gameplay duration (i.e. the speed of driving) can be affected by the emotional state of the opponent. In other words, it seems that the emotional awareness of the opponent increases the level of challenge in the game, which can consequently negatively affect the performance of players. One possible explanation is that the emojis representing the emotions of opponents on the screen are way too distracting. We argue, however, that being aware of one's emotions may result in complex social interactions between players that can, in turn, enhance the relationship between the player and the opponent. Another explanation behind the negative effect in the performance of players may be associated to higher levels of enjoyment as suggested by our findings: participants enjoyed seeing the other player's emotions. Consequently, it seems that although the performance of players is decreased due to higher levels of perceived competitiveness and cognitive effort required to process across-player emotional communications, players had a better experience and enjoyed the game more when they were aware of their opponent's emotional state.
	
	Another important question that remains to be addressed in a future study is whether the opponent's emotions directly affect the performance of a player. Is it, instead, the change of the players' own emotions which cause the difference in the behavior? As discussed earlier, the emotional states of the players were affected by changes in the emotional state of their opponents. Based on such findings there seems to be an association between the change in the player's performance and the change in her emotional state. We attribute this association to the complex emotional relations that are gradually built between the player and her opponent. This is in line with the reports of some participants claiming that they could not put up with happy opponents who were overtaking their car. Participants noted that even though they actively tried to cope with the displayed emotions of their opponents an additional emotional challenge was the control of their expressed emotions towards their opponents. Clearly, while more research is needed to shed light to the relationship between performance and emotion expression the current findings suggest that one affects the other and vice versa.
	
	An interesting behavior that we observed through the experiments was that some players were intentionally changing their facial expressions to affect the emotions of their opponent. For example, some players were intentionally laughing while passing the opponent's car to make their opponent angry. In other words, players were using the expression of their emotions as a medium to affect the emotions and behaviors of their opponents. Some of the participants reported that the ability to express emotions through facial expressions felt like a meta-game on top of the racing game. As multiplayer games use limited channels of direct interaction between players---e.g. chat boxes---we argue that emotion expression can be an additional channel for social interaction and gameplay dynamics in multi-player games. Presenting opponents' emotions to the player can open new opportunities for the design of gameplay dynamics and player experience, in which computer games can be equipped with emotion-aware interfaces that can be used to provide richer in-game interactivity across players.
	
	Although our findings show some notable benefits to gameplay when the opponent's emotions are presented to the player, a critical question arises: is the use of emojis the best way to express the emotions of one's opponent? We argue that different approaches of emotion expression can be used and tested across different game genres. Some game genres such as first-person shooters, for instance, require the full attention of the player at every moment of the gameplay. Thus, in such games the visual presentation of emotions on the screen may distract players, and hence auditory (e.g. sound effects) or haptic (e.g. vibration of controller) forms of emotion expression may prove better design choices. Some alternative and more direct forms of emotion expression can also be used in future studies. Instead of using emojis, for instance, the facial expression of the non player characters could alter in real-time based on the emotions of their opponents. Further research is required to test such options empirically and determine the impact of such design decisions across various game genres.
	
	Even though most of the participants agreed that presenting the emotions of opponents increased their sense of competition in the game, some players (especially those with low-level driving skills in car racing games) complained that the frequent changes of the emojis were disturbing, thereby, affecting their driving performance in a negative fashion. One possible reason for the frequent changes of the emojis is the natural behavior of a player that may frequently change his facial expression to distract the other players. We do not intend to address this problem if the reason for the frequent changes of the emojis is the natural behavior of the player. However, sometimes frequent changes happen because the facial expression is not active enough and it lies within the border of two states (e.g. between happy and neutral states) which makes it difficult for the algorithm to label the emotional expression. To address this challenge, one can assign a minimum lifetime for each manifested state thereby preventing over-frequent changes of the emojis. 
	A core challenge of the experimental protocol design was to find pairs of participants with similar skills in car racing games. For that purpose, we intentionally matched participants based on their self-reported driving skills in the car racing genre to cater for the balance of the games played and as a means to increase the number of changes in the ranking of players while playing. While such a match-up decision may differ from a natural car racing game setup that does not feature the profiling of players it is not uncommon that match-making occurs in multiplayer car racing games such as \emph{Forza Motorsport} (Microsoft Game Studios, 2005) and its \emph{Drivatar} system \cite{yannakakis2018artificial}.  
	
	It is clearly ambitious to claim that this paper offered a general conclusion on the validity of results for all game genres as the experiments were performed on a particular car racing game. We may argue, however, that the results obtained are generalizable within the car racing genre but also, to a degree, within games that feature a two-player competitive setting. Evidently, to test such a hypothesis we will need to conduct additional experiments in more game genres featuring large numbers of participants and more modalities of user input. 
	
	\section{Conclusion}
	
	This paper explored the impact of the expression of manifested emotions of an opponent on the emotions, and consequently the gameplay behavior, of players. We first showed that the emotional state of the opponent expressed by an emoji can attract the attention of players. Then, we showed that the expression of manifested emotions of an opponent can affect the game behavior of players in terms of speeding up, reducing speed, collisions, and gameplay duration. We also showed that the expression of manifested emotions of an opponent can affect the emotions of players in terms of increasing the rate of their emotional state changes. Our experiments concluded that the emotional awareness of the opponent can lead to an increase in the perceived competitiveness and enjoyment of the game.  
	
	The core findings of this paper suggest that the real-time recognition of opponents' emotions can affect and shape the future of game design and development. Processing human emotions via analyzing sensory data in games can result in affect-aware games, in which players' affective states are incorporated into the gameplay. In particular, computer games can be equipped with advanced interfaces that allow manifested emotions to be detected and used directly as means of player interactivity, beyond chat boxes and other popular means of reporting emotion that are used nowadays.

\bibliographystyle{IEEEtran}

\bibliography{refs}

\end{document}